\documentclass[prl,aps,floats,twocolumn,showpacs]{revtex4}
\usepackage[dvips]{epsfig}

\begin{document}

\def\sqr#1#2{{\vcenter{\hrule height.#2pt
   \hbox{\vrule width.#2pt height#1pt \kern#1pt
      \vrule width.#2pt}
   \hrule height.#2pt}}}
\def\square{{\mathchoice\sqr64\sqr64\sqr{3.0}3\sqr{3.0}3}}

\title{Vector boson mass generation without new fields}

\author{Bernd A. Berg}

\affiliation{Department of Physics, Florida State
             University, Tallahassee, FL 32306-4350, USA}

\date{March 7, 2012} % \date{\today } 

\begin{abstract}
Previously a model of only vector fields with a local U(1)$\,\otimes
\,$SU(2) symmetry was introduced for which one finds a massless U(1) 
photon and a massive SU(2) vector boson in the lattice regularization. 
Here it is shown that quantization of its classical continuum action 
leads to perturbative renormalization difficulties. But, non-perturbative 
Monte Carlo calculations favor the existence of a quantum continuum 
limit. 
\end{abstract}
\maketitle

% \section{Introduction}

Electromagnetic currents plus charged and neutral weak currents with 
the corresponding vector bosons ($\gamma,W^{\pm},Z^0$), are needed 
for a theory that accommodates simultaneously weak parity violation 
and electromagnetic parity conservation~\cite{Gl61}. Explicit breaking 
of gauge symmetry through massive vector bosons can be avoided by the 
Higgs mechanism~\cite{We67}, which leads to desired features such as 
perturbative renormalizability~\cite{tH71}. 

Nevertheless, the introduction of the Higgs particle into a theory
in which all matter fields are fermions with interactions mediated 
by vector bosons remains quite adhoc and the quadratic divergence of 
its self-energy causes a fine tuning problem~\cite{V81}. This provides 
an opening for proposing new physics solutions of which supersymmetry 
and string theory are most popular. Though new physics should emerge 
on the way to the Planck scale, there appear no strong reasons to 
expect experimental signals for it on the LHC energy scale. Occam's 
razor suggests to stay with fermions and vector bosons. 

Notably, the original arguments from the 1960s and early 
1970s rely on perturbation theory. Our non-perturbative understanding 
of quantum field theory (QFT) developed later. A milestone was Wilson's 
\cite{W74} formulation of lattice gauge theory (LGT) in 1974 and LGT 
Monte Carlo (MC) calculations started in earnest during the early 1980s 
after pioneering papers by Creutz and others~\cite{C80}. In recent 
years that the author suggested to address the problem of vector 
boson mass generation from scratch within the non-perturbative LGT 
framework~\cite{B10}. This appears to be of theoretical interest, 
independently of the question whether such a scenario is eventually
realized by nature or not.

Using Wilson's regularization, $V_{\mu}=\exp(ig_baB_{\mu})$, 
$B_{\mu}=\vec{\tau}\cdot\vec{b}_{\mu}/2$, where $\tau_i$, 
$i=1,2,3$ are the Pauli matrices, the SU(2) lattice action reads
\begin{eqnarray} \nonumber
  S_2 = \frac{\beta_b}{2} \sum_p {\rm Tr}\,V_p 
\end{eqnarray}
where the sum is over all plaquettes and $V_p$ are oriented products 
of SU(2) matrices around the plaquette loop. E.g., for a plaquette in 
the $\mu\nu$, $\mu\ne\nu$ plane $V_p$ becomes ($a$ lattice spacing, 
$x=na$ with $n$ an integer 4-vector):
\begin{eqnarray} \nonumber
  V_{\mu\nu}(x)&=&{\rm Tr}\,\left[V_{\mu}(x)\,V_{\nu}(x+\hat{\mu}a)
  \,V^{\dagger}_{\mu}(x+\hat{\nu}a)\,V^{\dagger}_{\nu}(x)\right]\,.
\end{eqnarray}
Due to non-perturbative effects there are no massless particles in the
spectrum of SU(2) LGT. The self interaction of the gauge fields creates
a spectrum of massive glueballs and one of them may be used to
set the mass scale. Coupled fermions are confined, while the leptons
are found as free particles. So, we need SU(2) LGT in a deconfined 
phase. This can be achieved by increasing the physical temperature 
\cite{Tc81}, but in a fundamental theory we have to stay at zero 
temperature.

The order parameter for the deconfining phase transition is the 
expectation value of the Polyakov loop. Polyakov loops are traces 
of products of SU(2) matrices along straight lines closed by periodic
boundary conditions. On a finite lattice one finds from confined to 
deconfined a transition between a single peak and a double peak 
distribution. A coupling of parallel SU(2) matrices,
$$ {\rm Re}\,{\rm Tr}\, \left[V_{\nu}(x+\hat{\mu}a)\,
                         V^{\dagger}_{\nu}(x)\right]\,, $$
is well suited to align Polyakov loops, but breaks SU(2) gauge
invariance. To rescue local U(1)$\,\otimes\,$SU(2) invariance of matter 
fields a concept of extended gauge invariance was introduced \cite{B10},
which requires to introduce additional vector fields.

Defining the U(1) field as $2\times 2$ matrix $U_{\mu}=\exp(ig_aa 
A_{\mu})$, $A_{\mu}=\tau_0\, a_{\mu}/2$ ($\tau_0$ unit matrix), we 
consider in the forthcoming the vector field lattice action 
\begin{equation} \label{S}
  S = \frac{\beta_a}{2} \sum_p {\rm Re}\,{\rm Tr}\,U_p
    + \frac{\beta_b}{2} \sum_p {\rm Tr}\,V_p 
    + \frac{\lambda}{2}\sum_{\mu\nu} S_{\mu\nu}^{\rm add}\,,
\end{equation}
where the third sum includes identical $\mu=\nu$ indices and
$$  S^{\rm add}_{\mu\nu}\ = {\rm Re}\,{\rm Tr}\,
   \left[U_{\mu}(x)\,V_{\nu}(x+\hat{\mu}a)\,U^{\dagger}_{\mu}
   (x+\hat{\nu}a)\,V^{\dagger}_{\nu}(x)\right]\,.               $$
The relations $\beta_a=1/g_a^2$ and $\beta_b=4/g_b^2$ define $\beta_a$ 
and $\beta_b$ of (\ref{S}) through the bare couplings constants and
$\lambda$ is a new free parameter.
Properties as function of $\lambda$ have been investigated by MC 
calculations~\cite{B10}. After fixing $\beta_a$ in the Coulomb phase 
of U(1) LGT and $\beta_b$ in the scaling region of confined SU(2) 
LGT, one finds for small $\lambda$ the same results 
as for $\lambda=0$: A massless U(1) photon and confined SU(2) gauge 
theory with a glueball spectrum. Increasing $\lambda$, a strong first 
order phase transition takes place. The U(1) photon survives the 
transition massless, while SU(2) is then in a deconfined phase with 
a massive vector boson triplet. Central questions are then about 
1.~Perturbative Renormalizability and 2.~Existence of a Quantum 
Continuum Limit. Both remained beyond the scope of~\cite{B10}. New 
insights are reported here after discussing a novel way to ensure
local U(1)$\,\otimes\,$SU(2) invariance.

Let us consider U(1)$\,\otimes\,$SU(2) field configurations $\{U'_{\mu}
(na)\}$ for which the SU(2) part is a gauge transformation of the zero 
field $B_{\mu}(na)\equiv 0$ and $\{V'_{\mu}(na)\}$ for which the U(1)
part is a gauge transformation of the zero field $A_{\mu}(na)\equiv 0$.
With $G\in$ U(1)$\,\otimes\,$SU(2) each set is mapped onto itself by
\begin{eqnarray} \label{aGU}
  U'_{\mu}(na) &\to& G(na) U'_{\mu}(na) G^{-1}(na+\hat{\mu}a)\,,
  \\ \label{aGV}
  V'_{\mu}(na) &\to& G(na) V'_{\mu}(na) G^{-1}(na+\hat{\mu}a)\,,
\end{eqnarray}
which we call {\it extended gauge transformations}. Replacing $U_{\mu}$ 
and $V_{\mu}$ by their primed versions, the action (\ref{S}) is invariant 
under these transformations. $S^{\rm add}_{\mu\nu}$ is the simplest 
example of Wilson loops that mix $U'_{\mu}$ and $V'_{\mu}$ matrices, 
which are now invariant operators.

The partition function of the model is
\begin{equation} \label{Zmodel} 
  Z\ =\ \int \prod_n\prod_{\mu=1}^4 dU'_{\mu}(na)\,dV'_{\mu}(na)\,e^S
\end{equation}
where the integrations are over $\{U'_{\mu}(na)\}$ and $\{V'_{\mu}(na)
\}$ defined above. They are easily implemented in a MC calculation. Let
us use the notation proper parts for the U(1) factor of the $U'_{\mu}$ 
and the SU(2) factor of the $V'_{\mu}$ matrices and the notation gauge
parts for the SU(2) factor of the $U'_{\mu}$ and the U(1) factor of 
the $V'_{\mu}$ matrices. The proper parts can be updated in the usual 
way. Specifically, a biased Metropolis-heatbath algorithm~\cite{Ba05}
was used in the simulations. Updates of the gauge parts transform all 
matrices emerging at a site $n$ according to
\begin{eqnarray} \label{GUV} 
  U'_{\mu}(na) \to G_2(na)\,U'_{\mu}(na)\,,
  V'_{\mu}(n)  \to G_1(na)\,V'_{\mu}(na)\,,
\end{eqnarray}
and all matrices on links ending at $n$ according to
\begin{eqnarray} \nonumber
  U_{\mu}(na-\hat{\mu}a)&\to& U_{\mu}(na-\hat{\mu}a)\,G_2^{-1}(na)\,,
  \\ \label{UGVG}
  V_{\mu}(na-\hat{\mu}a)&\to& V_{\mu}(na-\hat{\mu}a)\,G_1^{-1}(na)\,,
\end{eqnarray}
where $G_2$ and $G_1$ are, respectively, SU(2) and U(1) matrices
drawn with the group measure. These updates change \cite{err} the 
action (\ref{S}), so that a Metropolis algorithm will have an 
acceptance rate in the range $(0,1]$.

Changes of the action under (\ref{GUV}) can be undone by appropriate
updates of the proper parts of the matrices. Therefore, we can calculate 
operators that are invariant under extended gauge transformations in
any fixed gauge, i.e., omitting updates of the form~(\ref{GUV}). For 
MC calculations it is convenient to assign zero fields to the gauge
parts, which we call {\it proper gauge}. The purpose of extended 
gauge invariance is to allow an initial Lagrangian with local 
U(1)$\,\otimes\,$SU(2) invariance of matter fields without explicit 
breaking by a vector boson mass term.  Including matter fields our
Lagrangian in the classical continuum limit is (using Euclidean 
notation)
\begin{eqnarray} \label{L}
  &~& L =
    \overline{\psi}\left(i\gamma_{\mu}D^a_{\mu}-m\right)\psi 
  + \overline{\psi}\left(i\gamma_{\mu}D^b_{\mu}-m\right)\psi\,,
  \\ \nonumber &~& 
  - \frac{1}{2}{\rm Tr}\left(F^a_{\mu\nu}F^a_{\mu\nu}\right) 
  - \frac{1}{2}\left(F^b_{\mu\nu}F^b_{\mu\nu}\right)
  -\frac{\lambda}{4}\ {\rm Tr}\left(F^{\rm add}_{\mu\nu}
  F^{\rm add}_{\mu\nu}\right)\,.
\end{eqnarray}
Here $D^a_{\mu}=\partial_{\mu}+ig_aA'_{\mu}$ and $D^b_{\mu}=\partial_{
\mu}+ig_b B'_{\mu}$ are gauge covariant derivatives and the additional 
field tensor is
\begin{eqnarray} \label{Fadd}
  F^{\rm add}_{\mu\nu} &=& g_b\partial_{\mu}B'_{\nu} -
  g_a\partial_{\nu}A'_{\mu} + i\,g_ag_b\left[A'_{\mu},B'_{\nu}\right]\,.
\end{eqnarray}
The fermion field $\psi$ is assumed to be a doublet and the Lagrangian 
is invariant under the local U(1)$\,\otimes\,$SU(2) symmetry 
transformations $\psi\to G\,\psi$ with the continuum limit of 
extended gauge transformations for the vector fields being
$A'_{\mu}\to GA'_{\mu}G^{-1}+i(\partial_{\mu} G)G^{-1}/g_a$,
$B'_{\mu}\to GB'_{\mu}G^{-1}+i(\partial_{\mu} G)G^{-1}/g_b$~\cite{B10}.
This is the reason for the occurrence of two $\psi$ terms in the
Lagrangian.
\medskip

1. In the proper gauge, $B'_{\mu}\to B_{\mu}$, $A'_{\mu}\to A_{\mu}$,
one gets
\begin{equation} \label{Ladd}
  L^{\rm add}\ =\ 
  -\frac{\lambda\,g_a^2}{16}\,\left(\partial_{\mu}a_{\nu}\right)^2\
  -\frac{\lambda\,g_b^2}{16}\,\left(\partial_{\mu}b^i_{\nu}\right)^2\,.
\end{equation}
These pieces are found in U(1) and SU(2) effective Lagrangians, which 
are usually obtained by integrating over gauge transformations with 
a Gaussian weighting function. With the identifications $\xi=8/
(\lambda\,g_a^2)$ and $\xi=8/ (\lambda\,g_b^2)$, respectively, 
Eqn.~(9.56) and~(16.34) of~\cite{PS}. However, (16.43) comes here 
without the Faddeev-Popov ghost fields. Therefore \cite{PS}, the tree 
approximation is non-unitarity because of transitions to longitudinal 
modes, which require massive vector bosons while there is no explicit 
mass term in the Lagrangian~(\ref{L}), while the lattice regularization 
is unitary to the extent that one can prove reflection positivity. On 
the 1-loop level the vector boson self-energy is divergent, generating 
an infinite mass. These properties render the model ill-defined in 
conventional perturbation theory.

2. One may expect that the vector boson mass $am_W$ found in \cite{B10} 
is also non-perturbatively divergent. Then, the lattice regularization 
would not allow for a quantum continuum limit $am_W\to 0$. Instead, 
$am_W$ has to stay finite $am_W\ge am_{\min}>0$ in a smoothly connected 
range of couplings, eventually bounded by first order phase transitions. 
We investigate here the line
\begin{equation} \label{lambda}
  \beta_a=\lambda\,,~~\beta_b=2\lambda\,,~~\lambda\to\infty
\end{equation}
for which one could envision an approach to a quantum continuum limit 
in analogy to the behavior of asymptotically free non-Abelian gauge 
theories. The result is that our fits to the scenarios $am_W\to 
am_{\min}>0$ versus $am_W\to 0$ prefer the latter. % up to amazingly 
% large $\lambda$ values. %, though based on numerical results one 
% cannot exclude a turn around at even larger $\lambda$.

Our mass spectrum calculations were performed on lattices of size 
$N^3N_t$, $N_t\gg N$. For each value of $\lambda$ we have first to 
extrapolate the infinite volume limit $N\to\infty$ of $am_W(\lambda,N)$, 
denoted by $am_W(\lambda)=am_W(\lambda,\infty)$. Subsequently, we fit 
$am_W(\lambda)$ so that a $\lambda\to\infty$ extrapolation along
the line defined by (\ref{lambda}) can be performed. 

Our masses are deduced from correlation functions $c(t)$ of suitable 
trial operators by performing the usual two parameter cosh fits 
\begin{equation} \label{ct}
  c(t)\ =\ a_1\,\left[\,\exp(-am_Wt)+\exp(-am_W(N_t-t))\,\right]
\end{equation}
for a range of integers $0\le t_1\le t\le t_2$. Here the trial operators
\begin{equation} \label{Wboson}
  W_{i,\mu}(x)\ =\ -i\,{\rm Tr}\,\left[\tau_i\,W_{\mu}(x)\right]\,,
\end{equation}
are employed, where in slight deviation from \cite{B10} a U(1) phase is 
included, $W_{\mu}(x)=U^{\dagger}_{\mu}(x)\,V_{\mu}(x)$ (no summation 
over $\mu$). As the previously used operator, it becomes gauge invariant 
in combination with (static) fermion or boson fields, compare (6.20) of 
\cite{MM}. 

In addition correlations between trial operators for the U(1) photon 
and SU(2) glueball masses in the plaquette representations of the cubic 
group were calculated. Estimates of the U(1) photon mass are for all 
$\lambda$ consistent with zero, while there are no convincing signals 
in the glueball channels. This is similar to the results reported in 
\cite{B10} for one choice of coupling constant values.

\begin{table}[tb]
\caption{Mass estimates on $14^3N_t$ lattices and infinite volume 
extrapolations according to Eq.~(\ref{amwinfty}). \label{tab_amW}} 
\centering \medskip
\begin{tabular}{|c|c|c|c|c|}  \hline
% &\multicolumn{2}{c|}{$\lambda=0.4$}&\multicolumn{3}{c|}{$\lambda=0.9$}
% \\ \hline 
$\lambda$&$N_{\min}$&$am_W(\lambda,14)$&
                  $am_W(\lambda,\infty)$& $Q$  \\ \hline
   ~1.1  &6& 0.2659 (10) & 0.2658 (10) & 0.13 \\ \hline
   ~4.0  &4& 0.1245 (10) & 0.1180 (16) & 0.31 \\ \hline
   ~8.0  &4& 0.0905 (12) & 0.0876 (15) & 0.44 \\ \hline
   12.0  &4& 0.0740 (11) & 0.0719 (14) & 0.31 \\ \hline
   16.0  &4& 0.0675 (14) & 0.0653 (13) & 0.36 \\ \hline
   20.0  &4& 0.0597 (14) & 0.0552 (16) & 0.42 \\ \hline
   24.0  &4& 0.0523 (10) & 0.0500 (15) & 0.11 \\ \hline
   28.0  &6& 0.0519 (10) & 0.0487 (18) & 0.26 \\ \hline
   32.0  &4& 0.0480 (13) & 0.0448 (18) & 0.96 \\ \hline
\end{tabular} \end{table} % \vspace*{0.2cm}

\begin{figure}[tb] \begin{center} 
% Laptop/SU2U1Results/110901Results/fsamw.plt
\epsfig{figure=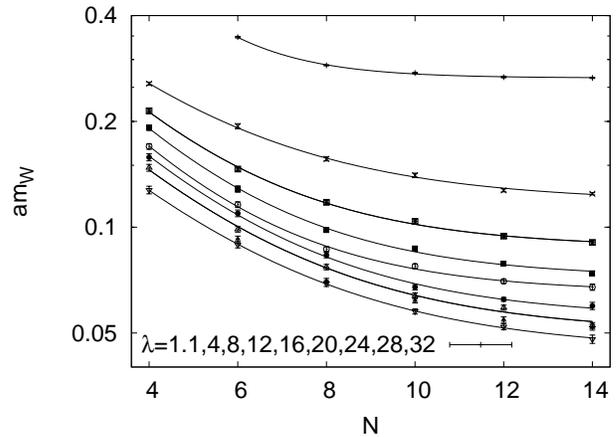,width=\columnwidth} % \vspace{-5mm}
\caption{Fits of $am_W(\lambda,N)$. The $\lambda$ values correspond 
in up to down order to the curves.  \label{fig_fsamw}} 
\end{center} \end{figure} % \vspace{-4mm}

Simulations were carried out on lattices of size $N=4,\,6,\,8,\,10,\,
12,\,14$ with $N_t$ in the range [48,96] and $\lambda$ values as 
given in table~\ref{tab_amW}. For each $\lambda$ the 3-parameter fit
\begin{equation} \label{amwinfty}
  am_W(\lambda,N)\ =\ am_W(\lambda,\infty)+a_2(\lambda)\,
                     \exp(-a_3(\lambda)N)
\end{equation}
was performed to derive an infinite volume estimate 
$am_W(\lambda,\infty)$. These fits are shown in Fig.~\ref{fig_fsamw}.
The extrapolations are collected in table~\ref{tab_amW} together 
with the goodness $Q$ of each fit and the estimates on our largest 
$14^3N_t$ lattices. Error bars are given in parenthesis and refer to 
the last digits in the number before. In two cases data from the 
smallest $4^3N_t$ lattice were omitted from the fit for consistency
reasons. We indicate with $N_{\min}$ the size of the smallest lattice 
included in the fit.

\begin{figure}[tb] \begin{center} 
% Laptop/SU2U1Results/110901Results/correlu1b20p0/cor20.plt.
\epsfig{figure=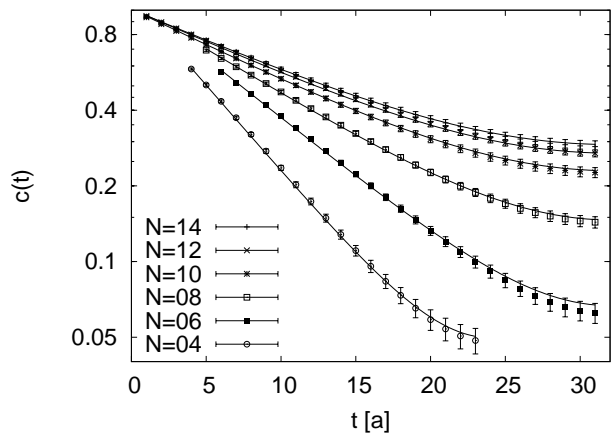,width=\columnwidth} %\vspace{-1mm}
\caption{Correlation functions for the $\lambda=20$ mass estimates.
The $N$ values correspond in up to down order to the curves.
\label{fig_cor20}} \end{center} \end{figure} % \vspace{-4mm}

To give an example of the numerical quality of the correlation functions 
(\ref{ct}) we depict them in Fig.~\ref{fig_cor20} for our lattices at 
$\lambda=20$. In stark contrast to the noise one encounters for glueball 
correlations in pure lattice gauge theories, these are beautiful strong 
correlations. One can easily follow them over more than 30 lattice 
spacing, though the estimates for our $14^3N_t$ lattices are already 
rather time consuming. Relying on a statistics of 500,000 sweeps, they 
run with the present single processor code one week on an Intel i7 CPU. 
% As small $am_W$ values are difficult to distinguish from one another 
% within their statistical noise, calculations aiming at even larger 
% $\lambda$ values would have to move to a different computational scale 
% like parallel processing on supercomputers.

\begin{table}[tb] % /Laptop/SU2U1Results/110901Results/u1bfits/readme.
\caption{Fits of the $am_W(\lambda)=am_W(\lambda,\infty)$ values of 
table~\ref{tab_amW}. The first column gives the number of parameters,
the last column the goodness of the fit. \label{tab_fits}} 
\centering \medskip
\begin{tabular}{|c|c|c|c|} \hline
par \#& function    &$am_W(\infty)$& $Q$  \\ \hline
2 &$ f_1(\lambda)=a_1\exp(-a_2\lambda)               $&0           &            0      \\ \hline
3 &$ f_2(\lambda)=a_0+f_1(\lambda)                   $&0.05783 (66)&            0      \\ \hline
3 &$ f_3(\lambda)=a_1\lambda^{-a_2}\exp(-a_3\lambda) $&0           &$1.2\times 10^{-6}$\\ \hline
4 &$ f_4(\lambda)=a_0+f_3(\lambda)                   $&0.03 (18)   &$1.1\times 10^{-4}$\\ \hline
4 &$ f_5(\lambda)=f_3(\lambda)\,(1+a_4/\lambda)      $&0           & 0.26 \\ \hline
5 &$ f_6(\lambda)=a_0+f_4(\lambda)                   $&0.02 (11)   & 0.20 \\ \hline
3 &$ f_7(\lambda)=a_1\lambda^{-a_2}(1+a_4/\lambda)   $&0           & 0.036\\ \hline
4 &$ f_8(\lambda)=a_0+f_7(\lambda)                   $&$-$0.1 (1.1)& 0.065\\ \hline
2 &$ f_9(\lambda)=a_1\lambda^{-a_2}                  $&0           &$4.7\times 10^{-15}$\\ \hline
10&$ f_{10}(\lambda)=a_0+f_9(\lambda)                $&0.0241 (26) &$2.0\times 10^{-4}$\\ \hline
\end{tabular} \end{table} % \vspace*{0.2cm}

We are now prepared to discuss the $\lambda\to\infty$ behavior of the
$am_W(\lambda)=am_W(\lambda,\infty)$ values of table~\ref{tab_amW}.
A number of fits, either enforcing $am_W(\lambda)\to 0$ for $\lambda
\to\infty$ or allowing for a free parameter $am_W(\infty)$ were tried 
and are compiled in table~\ref{tab_fits}. The fit forms $f_1(\lambda)$ 
to $f_4(\lambda)$ are in disagreement with the data as signaled by very 
small $Q$ values in the last column. A zero means that $Q$ is so small 
that the precise value cannot be tracked within our rounding errors of 
about $10^{-20}$. 
%
% /Laptop/SU2U1Results/110901Results/u1bfits/MWlambda.plt.
\begin{figure}[tb] \begin{center} 
\epsfig{figure=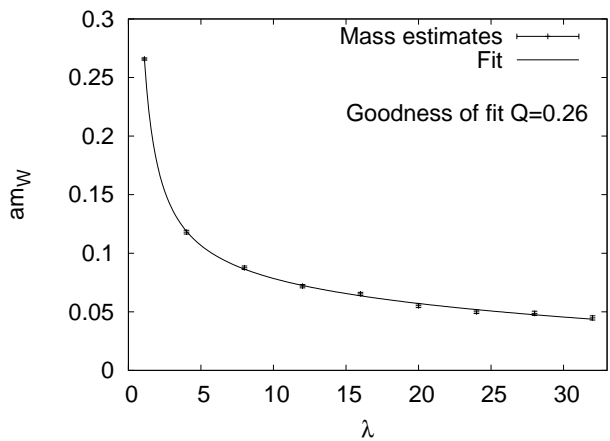,width=\columnwidth} %\vspace{-1mm}
\caption{Best fit of $am_W(\lambda)$. 
\vspace{-4mm} \label{fig_amw} } \end{center} \end{figure}
Most convincing is our 4-parameter fit
\begin{equation} \label{amw}
  a\,m_W(\lambda)\,=\,f_5(\lambda)\, =\, a_1\,\lambda^{a_2}\,
      \exp(a_3\lambda)\,(1+a_4/\lambda)
\end{equation}
with a well acceptable goodness of fit $Q=0.26$. The values of its 
parameters are $a_1=0.111\,(47)$, $a_2=-0.16\,(18)$, $a_3=-0.0139\,
(75)$, $a_4=1.6\,\left(\matrix{+2.1\cr -0.6}\right)$. When adding 
$am_W(\infty)$ as 5th parameter $a_0$, the function $f_6(\lambda)$ 
is obtained for which the goodness of fit goes slightly down, 
indicating that we are overfitting, and the estimated $a_0=
am_W(\infty)$ includes zero.

The functional form (\ref{amw}) has similarities with the asymptotic 
freedom behavior in LGT. As the estimated value of the
factor $a_3$ in the exponent is small, one wonders whether a power
law alone suffices to describe $m_W(\lambda)$. The functional forms 
$f_7(\lambda)$ to $f_{10}(\lambda)$ test this. While the last two
of them are bad, this is less obvious for $f_7(\lambda)$ and 
$f_8(\lambda)$. Recall, under the assumption that the form of a 
fit is correct, $Q$ is the probability for the discrepancy between 
the fit and the data.

The presented MC calculations indicate divergence of the correlation 
length $\xi/a\to\infty$, $\xi=m_W^{-1}$ for $\lambda\to\infty$,
contradicting perturbation theory and supporting a quantum continuum 
limit. Of course, we cannot exclude that for larger systems and 
coupling constants the behavior may turn around and support $m_W(
\lambda)\to m_{\rm min}>0$. While this is correct, the perturbative 
approach is in essence vulnerable to similar criticism. Though it 
served us well, there is no proof that perturbation theory describes 
the true nature of a QFT. 

Besides moving on to new horizons, we should perhaps keep an open 
mind for the third logical possibility that a deeper understanding 
of conventional QFT could unveil new models with massive vector
bosons. The unexpected numerical results of this paper provide 
no answers, but indicate that it is worthwhile to continue this 
line of work. 
\bigskip

\acknowledgments 
This work was in part supported by the DOE grant DE-FG02-97ER41022.
Constructive criticism by PRL referee B led to relevant improvements
of the paper. Some of the computer programs used rely on collaborations 
with Alexei Bazavov.

\end{document}